\documentclass[twocolumn,showpacs,preprintnumbers,superscriptaddress, 
amsmath,amssymb,prl,aps,floatfix]{revtex4}

\usepackage{epsfig}
\usepackage{amsmath}
\usepackage{amssymb}
\usepackage{bm}
\usepackage{url}
\usepackage{graphicx}

\begin{document}

\title{Direct High-Power Laser Acceleration of Ions for Medical 
Applications}

\author{Yousef I. Salamin}
\email{ysalamin@aus.edu}
\affiliation{Max-Planck-Institut f\"{u}r Kernphysik, Saupfercheckweg 1, 
69117 Heidelberg, Germany}
\affiliation{Physics Department, American University of Sharjah, POB 
26666, Sharjah, United Arab Emirates}
\author{Zolt\'{a}n Harman}
\email{harman@mpi-hd.mpg.de}
\author{Christoph H. Keitel}
\email{keitel@mpi-hd.mpg.de}
\affiliation{Max-Planck-Institut f\"{u}r Kernphysik, Saupfercheckweg 1, 
69117 Heidelberg, Germany}

\begin{abstract}

Theoretical investigations show that linearly and radially polarized 
multiterawatt and petawatt laser beams, focused to subwavelength waist 
radii, can directly accelerate protons and carbon nuclei, over 
micron-size distances, to the energies required for hadron cancer 
therapy. Ions accelerated by radially polarized lasers have generally a 
more favorable energy spread than those accelerated by linearly 
polarized lasers of the same intensity.

\end{abstract}


\maketitle

Protons and heavier ions are now being used to treat cancer at a number 
of places around the world \cite{fritzler}. Ion lithography schemes seem 
to be heading for practical application \cite{lithog} and fusion 
research continues to attract considerable attention and to gain in 
importance \cite{roth1}. In addition, considerable effort is being 
devoted to research into the fundamental forces of nature, the 
initiation of nuclear reactions and into schemes to treat radioactive 
waste \cite{mckenna}. In these applications, conventional accelerators 
(synchrotrons, cyclotrons and linacs) are employed which are large and 
expensive to build and operate.

To produce and accelerate ions, current plasma-based research focuses on 
the use of thin foils irradiated by femtosecond laser pulses of 
intensity $>10^{18}$ W/cm$^2$ \cite{badziak}. Typically a laser pulse is 
incident on the thin foil giving rise to an overdense plasma from which 
the electrons get accelerated, form a dense sheath on the opposite side 
and generate a quasistatic electric field of strength in excess of 
$10^{12}$ V/m. This superstrong field accelerates the ions to tens of 
MeV over a distance in the $\mu$m range \cite{mackinnon}. Recent work 
\cite{maksim} has shown that proton beams produced by this method of 
target normal sheath acceleration (TNSA) may be improved in energy and 
beam quality by the use of foils less than 1 $\mu$m in thickness 
\cite{neely}. In earlier experiments, employing thicker foils, a small 
fraction of the energy got converted to proton energy. Furthermore, the 
protons had energy spreads reaching 100\%.

In hadron therapy \cite{bulanov,hicat,svensson}, for example, the ions 
are required to have kinetic energies $K=20-250$ MeV/nucleon (H$^+$ and 
He$^{2+}$) and $K=85-430$ MeV/nucleon (C$^{6+}$ and O$^{8+}$). The 
ions also ought to have an energy spread $\Delta K_f/\bar{K}_f < 1\%$ so 
that they may be focused on the tumor while sparing the neighboring 
healthy tissue. A beam of rectangular cross section is also desirable 
\cite{svensson}.

In this Letter we study direct laser acceleration configurations of 
protons and bare carbon nuclei. The aim is to make predictions regarding 
the optimum conditions that would lead to the ion energies of interest 
to hadron therapy. A source for the ions may be a dedicated electron 
beam ion trap/source (EBIT/EBIS)~\cite{crespo} from which they can be 
extracted in a well-defined fully ionized charge state or an ensemble of 
fully stripped ions produced by laser-solid interaction. We consider a 
situation in which the source has been tailored on a nanoscale 
\cite{bulanov,roth}. We study the dynamics of an ensemble of $N$ 
particles having normally distributed kinetic energies, using the 
single-particle relativistic Lorentz-Newton equations. Our calculations 
show that laser pulses generated by 0.1$-$10 PW laser systems accelerate 
ions directly to energies in the ranges required for hadron therapy, 
provided the laser beams are focused to subwavelength waist radii. The 
accelerated ions turn out to have high beam quality and rectangular and 
circular cross sections reflecting symmetry of the field and shape of 
the initial ionic distribution.

Key to generating the ultrastrong accelerating fields is focusing the 
laser beam to a subwavelength waist radius. According to recent 
experiments, a linearly polarized beam of wavelength $\lambda$ may be 
focused to a spot of size $(0.26\lambda)^2$, where the spot size is the 
area enclosed by a contour at which the beam intensity falls to one half 
its peak value. A radially polarized beam may be focused to the 
substantially smaller spot size of $(0.16\lambda)^2$ \cite{quabis1}.

Upon tight focusing, a linearly polarized laser beam develops five field 
components. The nonvanishing field components of a beam polarized along 
the $x$ axis, propagating along the $z$ axis, of wavelength $\lambda$ 
and frequency $\omega$, are given in~\cite{sal-apb}, using the familiar 
Gaussian-beam parameters (waist radius $w_0$, Rayleigh length $z_r=\pi 
w_0^2/\lambda$ and diffraction angle $\varepsilon=w_0/z_r$). The laser 
power expression may be given, to order $\varepsilon^{10}$, by
\begin{eqnarray}\label{pow-linear}
  \label{pow} P_l &=& \frac{\pi w_0^2}{4}\frac{E_{0l}^2}{c\mu_0}
  \left[1 + \left(\frac{\varepsilon}{2}\right)^2 + 2
\left(\frac{\varepsilon}{2}\right)^4 + 6
\left(\frac{\varepsilon}{2}\right)^6 \right. \nonumber \\
& &\left. + \frac{45}{2}
\left(\frac{\varepsilon}{2}\right)^8 + \frac{195}{2}
\left(\frac{\varepsilon}{2}\right)^{10} \right],
\end{eqnarray}
where $\mu_0$ is the permeability of free space, $E_{0l}$ is the 
electric field amplitude, $c$ is the speed of light in vacuum and the 
subscript $l$ stands for linearly polarized. Note that 
$E_{0l}\propto\sqrt{P_l}$ and that the leading term in $E_{0l}$ is 
inversely proportional to $w_0$.

\begin{figure}[t]
\includegraphics[width=7.2cm,height=!]{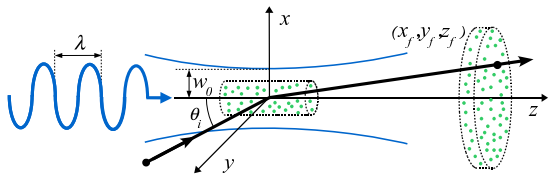}
\caption{
(Color online) Schematic geometry of the acceleration scenario. The 
initial coordinates of the ions are randomly distributed in a cylinder 
which models the interaction regime of the ion beam and the laser field. 
The ejected ions then form a beam of circular cross section when the 
ions are accelerated by radially polarized laser pulses. See text for 
further details.
}
\label{fig1}
\end{figure}

Dynamics of a particle of mass $M$ and charge $Q$ in electric and 
magnetic fields is governed by the equations of motion
\begin{equation}\label{motion}
    \frac{d\bm{p}}{dt}=Q[\bm{E}+c\bm{\beta}\times\bm{B}];\quad \frac{d{\cal
    E}}{dt}=Qc\bm{\beta}\cdot\bm{E},
\end{equation}
in which the energy and momentum of the particle are given by ${\cal 
E}=\gamma Mc^2$ and $\bm{p}=\gamma Mc\bm{\beta}$, respectively, with 
$\bm{\beta}$ its velocity scaled by $c$, and $\gamma=(1-\beta^2)^{-1/2}$ 
its Lorentz factor. Integrating Eqs. (\ref{motion}) numerically one 
obtains $\bm{\beta}$ and, hence, $\gamma_f$ at a later time $t_f$ taken 
equal to many laser field cycles. The final kinetic energy is
$K_f=(\gamma_f-1)Mc^2$.

To model the interaction region of the ion beam and the laser we 
consider an ensemble of $N=5000$ particles initially randomly 
distributed in a cylinder of radius $R_c=10$~nm and length $L_c=100$~nm 
oriented along the $z$ axis and centered on the origin (see Fig. 
\ref{fig1}). In order to simulate a realistic ion beam extracted from 
some ion source, the particles will be assumed to possess Gaussian 
distributed random initial kinetic energies with a mean value 
$\bar{K}=10$~keV and a spread $\Delta K=10$~eV. Initial direction of 
motion of all particles will be in the $xz$ plane and at 
$\theta_i=10^\circ$ relative to the pulse propagation direction.

\begin{figure}[t]
\includegraphics[width=8.0cm,height=5.2cm]{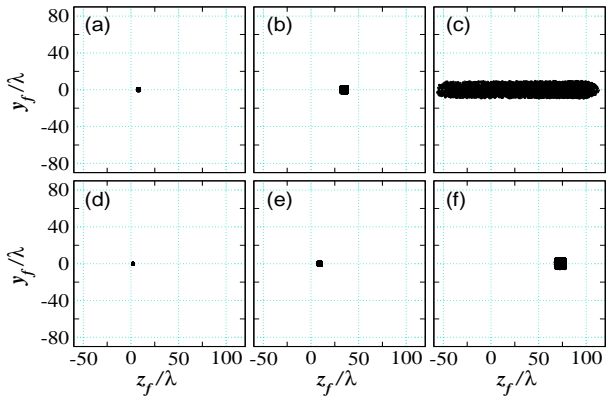}
\caption{
(Color online) Coordinates $(z_f,y_f)$ of 5000 particles at the end of 
their trajectories after interaction with linearly polarized light. 
(a)-(c) are for protons $^1$H$^+$ and (d)-(f) are for carbon nuclei 
$^{12}$C$^{6+}$. The parameters are: $\lambda=1054$ nm, 
$w_0=0.5\lambda$. The equations have been 
integrated over a time interval equivalent to 1000 laser field cycles, 
approximately amounting to 3.516~ps. For further data see 
Table~\ref{table1}.
}
\label{fig2}
\end{figure}

\begin{table}[b]
\caption{
Data related to Fig. \ref{fig2}. Laser power, average particle final 
$x$ coordinate $\bar{x}_f$, average final kinetic energy $\bar{K}_l$ and 
relative kinetic energy spread $\Delta K_l/\bar{K}_l$ at the end of the 
trajectories of protons, (a)-(c), and carbon nuclei, (d)-(f). The ions 
are interacting with linearly ($l$) polarized light. \label{table1}
}
\begin{ruledtabular}
\begin{tabular}{ccccc}
~ & Power & $\bar{x}_f$ & $\bar{K}_l$ & $\Delta K_l/\bar{K}_l$\\
~ & [PW]  & [$\lambda$] & [MeV/nucleon] & [\%]\\
\hline
(a) & 0.1  & $89.6 \pm 0.6$ & $3.77  \pm  0.05$ & $1.3$\\
(b) & 1    & $283.0  \pm 1.7$ & $37.39 \pm  0.45$ & $1.2$\\
(c) &10    & $750.6  \pm 41.1$  & $416.5 \pm  24.7$ & $5.9$\\
(d) & 0.1  & $44.9 \pm 0.3$ & $0.94  \pm  0.01$ & $1.2$\\
(e) & 1    & $141.3  \pm 0.9$ & $9.32  \pm  0.11$ & $1.2$\\
(f) &10    & $434.3  \pm 2.5$ & $89.9\pm  0.9$  & $1.0$\\
\end{tabular}
\end{ruledtabular}
\end{table}

All $\bm{E}$ components, together with the $\bm{v}\times\bm{B}$ force, 
work to accelerate a particle and deflect it to varying degrees from its 
initial direction of motion \cite{sal-apb}. A particle makes the longest 
excursion along the $x$ axis due to $E_x$ being the strongest 
accelerating component, smaller excursions occur due to $E_y$ and $E_z$. 
An ensemble of such particles forms a beam of rectangular cross section 
in the $yz$ plane, reflecting symmetry of the laser field. This is 
demonstrated in Fig. \ref{fig2} for interaction with 0.1, 1 and 10 PW 
laser beams. The ion beam cross section increases with increasing laser 
power ($E_{0l}\propto\sqrt{P_l}$). The transverse proton beam divergence 
is larger than for carbon. Energy gain by protons is also greater than 
the gain per nucleon of carbon, due to the proton's larger charge to 
mass ratio. Further data not estimable from Fig. \ref{fig2} are 
collected in Table \ref{table1}. Note that the energies fall within the 
domain required for hadron therapy and their spread is close to what 
would be suitable for such a purpose. Our calculations also show that 
the relative energy spread tends to increase approximately linearly with 
the volume of the initial ionic distribution. However, the input (and 
output) particle beam may be collimated using externally applied 
electromagnetic fields (see, e.g. Ref.~\cite{verschl}).

\begin{figure}[t]
\includegraphics[width=8.0cm,height=8.0cm]{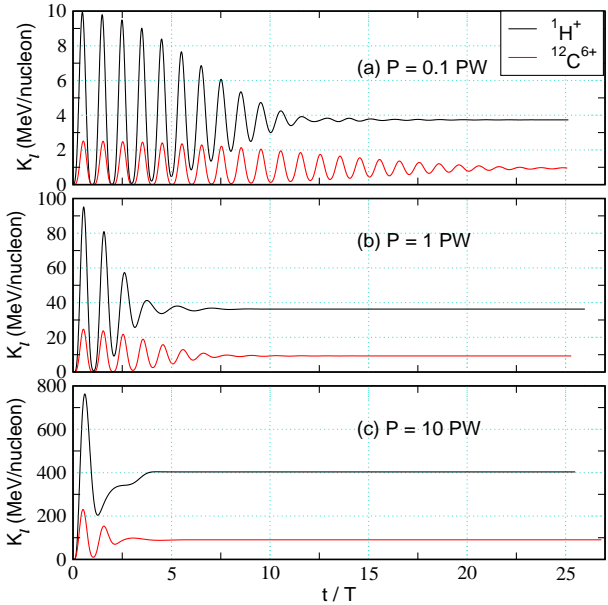}
\caption{
(Color online) Kinetic energy of a typical particle (out of the 
ensembles of Fig. \ref{fig2}) during interaction with linearly polarized 
light. In the figure, T is one laser field period.
}
\label{fig3}
\end{figure}

For realization, a short pulse consisting of only a small number of 
field cycles would be needed. This conclusion is elucidated by showing, 
in Fig. \ref{fig3} for a typical member of the ensemble, the particle 
kinetic energy as a function of the number of interaction cycles. The 
number of actual interaction field cycles is small and decreases with 
increasing power, since ions accelerated to higher velocities leave the 
focal region faster. Laser-to-particle energy conversion occurs in the 
form of a few violent impulses. Note also that, due to its lower 
velocity, a carbon ion interacts with more field cycles than a proton.

Estimates may be obtained for the transverse and longitudinal emittances 
$\epsilon_T$ and $\epsilon_L$, respectively, of the particle beam. Using 
ensemble averages of the exit particle coordinates and kinetic energies, 
one calculates $\epsilon_T < 10 \pi$ mm~mrad and $\epsilon_L < 
4\times10^{-5}$ eV~s for protons (case of Fig. 2(c)).  A comparison of 
these figures with results recently reported for ion beams produced from 
ultrathin solid foils irradiated by multiterawatt and petawatt linearly 
polarized laser beams \cite{roth,roth2} reveals that a good quality 
proton beam may be obtained by direct laser acceleration. Estimates for 
carbon yield $\epsilon_T < 0.3\pi$ mm~mrad and $\epsilon_L < 
9\times10^{-6}$ eV~s (case of Fig. 2(f)). The remaining cases in Fig. 2 
have better emittances.

A Gaussian beam of linear polarization is relatively easy to generate in 
the laboratory, while the same thing may not be said about a beam of the 
radially polarized variety. The electric field of a radially polarized 
focused (axicon) laser beam has two components, radial $E_r$ and axial 
$E_z$, with propagation along the $z$ axis. However, only one 
azimuthally polarized magnetic field component, $B_\theta$, exists. 
$E_z$ works efficiently to accelerate the particles, while $E_r$ and 
$B_\theta$ help to limit their diffraction \cite{zubairy,birula}. 
Full expressions for the axicon field components are given in~\cite{sal-njp}.
The power of the axicon beam, to order $\varepsilon^{10}$, reads
\begin{eqnarray}\label{pow-radial}
  P_r &=& \frac{\pi w_0^2}{2}\frac{E_{0r}^2}{c\mu_0}
  \left(\frac{\varepsilon}{2}\right)^2
  \left[1 + 3 \left(\frac{\varepsilon}{2}\right)^2+
    9 \left(\frac{\varepsilon}{2}\right)^4
\right.\nonumber\\
& &\left.
+30 \left(\frac{\varepsilon}{2}\right)^6
+ \frac{225}{2} \left(\frac{\varepsilon}{2}\right)^8\right].
\end{eqnarray}
Note that the amplitude $E_{0r}\propto\sqrt{P_r}$. On the other hand, 
when the definition of $\varepsilon$ is used, one finds that the leading 
term in $E_{0r}$ is independent of $w_0$. While $E_{0l}$ has a peak 
value beyond which it falls asymptotically to zero, with increasing 
$w_0$, $E_{0r}$ increases to an asymptotic value.

\begin{figure}[t]
\includegraphics[width=8.0cm,height=5.2cm]{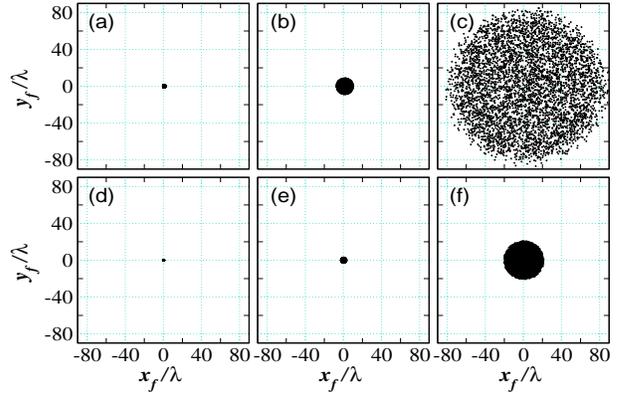}
\caption{
(Color online) Same as in Fig. \ref{fig2}, for particles accelerated by 
radially polarized light. For further data see Table~\ref{table2}.
}
\label{fig4}
\end{figure}

\begin{table}[b]
\caption{
Data related to Fig. \ref{fig4}. Same as Table \ref{table1}, for
ions interacting with radially ($r$) polarized light. \label{table2}
}
\begin{ruledtabular}
\begin{tabular}{ccccc}
~ & Power & $\bar{z}_f$  & $\bar{K}_r$  & $\Delta K_r/\bar{K}_r$\\
~ & [PW]  & [$\lambda$] & [MeV/nucleon]& [\%]\\
\hline
(a) & 0.1  & $104.7\pm0.5$   & $4.26\pm0.03$  & $0.7$\\
(b) & 1    & $430.8\pm2.3$   & $45.7\pm0.4$   & $0.8$\\
(c) &10    & $3347.5\pm89.6$ & $532.8\pm13.3$ & $2.5$\\
(d) & 0.1  & $48.5\pm0.2$    & $1.00\pm0.01$  & $1$\\
(e) & 1    & $173.7\pm0.8$   & $10.4\pm0.1$   & $0.8$\\
(f) &10    & $876.1\pm5.3$   & $122.0\pm1.0$  & $0.8$\\
\end{tabular}
\end{ruledtabular}
\end{table}

\begin{figure}[t]
\includegraphics[width=8.0cm,height=8.0cm]{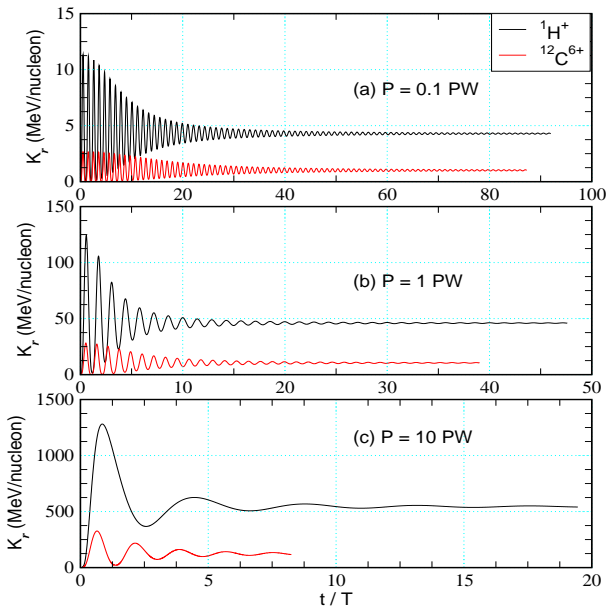}
\caption{
(Color online) Kinetic energy of a typical particle (out of the 
ensembles of Fig. \ref{fig4}) during interaction with radially polarized 
light.
}
\label{fig5}
\end{figure}

Figs. \ref{fig4} and \ref{fig5} are similar to Figs. \ref{fig2} and 
\ref{fig3}, respectively, albeit for particle acceleration by radially 
polarized laser fields. From Fig. \ref{fig4} one sees that the particle 
beam cross section in the $xy$ plane is almost circular, reflecting 
symmetry of the initial distribution. Recall that the radial electric 
field component $E_r$ vanishes on the beam axis and its magnitude at a 
typical particle trajectory is smaller than the transverse field in the 
linearly polarized case. In addition, $E_r$ plays a confining role 
during the half-cycle in which it is radially inward and tends to cause 
dispersion during the following half cycle. The fact that most of the 
beam power is concentrated in $E_z$, due to tight focusing, is also 
responsible for the extra energy gain from the radially polarized beam 
compared to that obtained from a linearly polarized one. Comparison of 
Figs. \ref{fig5} and \ref{fig3} also shows that a particle, since it is 
confined to dominantly axial motion, interacts with many more radially 
polarized cycles than it does with linearly polarized ones.

To assess the beam qualities, estimates show that, for protons, 
$\epsilon_T < 3 \pi$ mm mrad and $\epsilon_L < 4\times10^{-5}$ eV s 
(case of Fig. 4(c)), while for carbon $\epsilon_T < \pi$ mm mrad and 
$\epsilon_L < 4\times10^{-6}$ eV s (case of Fig. 4(f)). The remaining 
cases in Fig. 4 have better emittances. These results are several orders 
of magnitude lower than their counterparts in conventional accelerators. 
They are also comparable to, and sometimes even better than, the 
corresponding figures for ion beams produced from ultrathin foils 
\cite{roth,roth2}.

The average kinetic energies gained from interaction with radially 
polarized laser beams are somewhat better than from interaction with the 
linearly polarized ones (compare Tables \ref{table1} and \ref{table2}). 
These energies, with their low spread, cover the domain of application 
in hadron therapy. Note that the initial ensemble volume corresponds to 
an ion density $n_i\sim10^{20}~cm^{-3}$. Estimations of the energies 
transferred via space-charge effects suggest that those are negligible 
for the parameters applied. One could, in principle, start out with 
ensemble parameters that would lower this density down to 
$n_i\sim10^{17}~cm^{-3}$ and still keep the accelerated beam quality 
within the limits of utility in ion therapy. Unfortunately, these 
densities are many orders of magnitude higher than what is available 
today from conventional ion sources. Further design improvements on such 
machines need to be made before they may be used as ion sources for 
direct acceleration. An alternative source could possibly be a solid 
target of micro- or nanoscale thickness, perhaps blown off by a laser 
pulse, as in the TNSA mechanism \cite{maksim,neely} and the laser-piston 
regime \cite{esirkepov}, followed at the appropriate time delay by an 
accelerating pulse of the type discussed in this Letter.

In conclusion, direct laser acceleration of ions is put forward as an 
appealing alternative for utilization in cancer therapy. Radially 
polarized beams of the required power and intensity are not currently 
available. Employing an existing linearly polarized beam, on the other 
hand, may call for a focusing mechanism to bring the ion beam spreading 
down and render it useful in applications.

Z.H. acknowledges insightful conversations with Jos\'e R. Crespo 
L\'opez-Urrutia.

\end{document}